\documentclass{aa}
\usepackage{psfig}
\usepackage{times}
\begin{document}
%
%
\headnote{Letter to the Editor}
\title{The composition of circumstellar gas and dust in 51~Oph 
\thanks{Based on observations with ISO, an ESA project with instruments
        funded by ESA Member States (especially the PI countries: France,
        Germany, the Netherlands and the United Kingdom) and with the
        participation of ISAS and NASA.}}

\author{M.E. van den Ancker\inst{1,2} \and G. Meeus\inst{3} \and 
        J. Cami\inst{2,4} \and L.B.F.M. Waters\inst{2,3} \and C. Waelkens\inst{3}}
\institute{
Harvard-Smithsonian Center for Astrophysics, 60 Garden Street, MS 42, 
 Cambridge  MA 02138, USA \and
Astronomical Institute, University of Amsterdam, 
 Kruislaan 403, 1098 SJ  Amsterdam, The Netherlands \and
Instituut voor Sterrenkunde, Katholieke Universiteit Leuven, Celestijnenlaan 
 200B, B-3001 Heverlee, Belgium \and
SRON, P.O. Box 800, 9700 AV  Groningen, The Netherlands
}

\offprints{M.E. van den Ancker (mvandenancker@cfa.harvard.edu)}
\date{Received <date>; accepted <date>} 

\abstract{
We analyze ISO archive data of the nearby bright emission-line star
51~Oph, previously classified as a proto-planetary system similar to
$\beta$~Pic.  The infrared spectrum reveals the presence of gas-phase
emission bands of hot ($\sim$ 850~K) CO, CO$_2$, H$_2$O and NO.  In
addition to this, partially crystalline silicate dust is present.  The
solid-state bands and the energy distribution are indicative of dust
that has formed recently, rather than of debris dust.  The presence of
hot molecular gas and the composition of the circumstellar dust are
highly unusual for a young star and are reminiscent of what is found
around evolved (AGB) stars, although we exclude the possibility of
51~Oph belonging to this group.  We suggest several explanations for
the nature of 51~Oph, including a recent episode of mass loss from a
Be star, and the recent destruction of a planet-sized body around a
young star.
}

\maketitle

\section{Introduction}
51~Ophiuchi (HR~6519) is a bright ($V$ = 4.78), nearby ($d$ = 131~pc)
emission-line star. Although the star has traditionally been
classified as B9.5Ve (Buscombe 1963; Slettebak 1982; Jaschek \&
Jaschek 1992; Dunkin et al. 1997), recent work by Gray \& Corbally
(1998) show 51~Oph to be of spectral type A0II--IIIe, more in line
with its position to the right of the main-sequence in the
Hertzsprung-Russell diagram (van den Ancker et al. 1998). For the
current generation of ground-based telescopes, 51~Oph is a
point-source at visual to mid-infrared wavelengths (Merkle et
al. 1990; Lagage \& Pantin 1994).

51~Oph exhibits a normal UV and visual energy distribution, but IRAS
and subsequent ground-based follow-up observations revealed an
unusually large infrared excess longward of 2~$\mu$m, interpreted as
being due to hot (up to 1000~K) circumstellar dust (Cot\'e \& Waters
1987; Waters et al. 1988). The peculiarity of the energy distribution
of 51~Oph when compared to those of Be, Ae and A-shell stars led
Waters et al. (1988) to suggest that it may be a candidate
proto-planetary disk system, similar to $\beta$ Pictoris. This
suggestion was further strengthened by Grady \& Silvis (1993), who
found evidence for the presence of variable columns of accreting gas,
similar to those found around $\beta$~Pic.

The presence of dust in the 51~Oph system was unambiguously
established in 1993, when Fajardo-Acosta et al. detected a prominent
10~$\mu$m silicate feature in emission. Their results were
subsequently confirmed from the ground by Lynch et al. (1994), Walker
\& Butner (1995) and Sylvester et al. (1996) and by Waelkens et
al. (1996) using the {\it Infrared Space Observatory} (ISO; Kessler et
al. 1996).  
Lecavelier des Etangs et al. (1997) found cold neutral C
($N$ = $5 \times 10^{13}$~cm$^{-2}$, $T$ = 20~K) in 51~Oph, yielding a
C\,{\sc i}/dust ratio similar to that derived for $\beta$~Pic.  Since
C\,{\sc i} has a very short lifetime, it must be continuously
replenished, providing evidence for the existence of evaporating
bodies in 51~Oph, similar to those inferred around $\beta$~Pic.
\begin{figure*}[t]
\centerline{\psfig{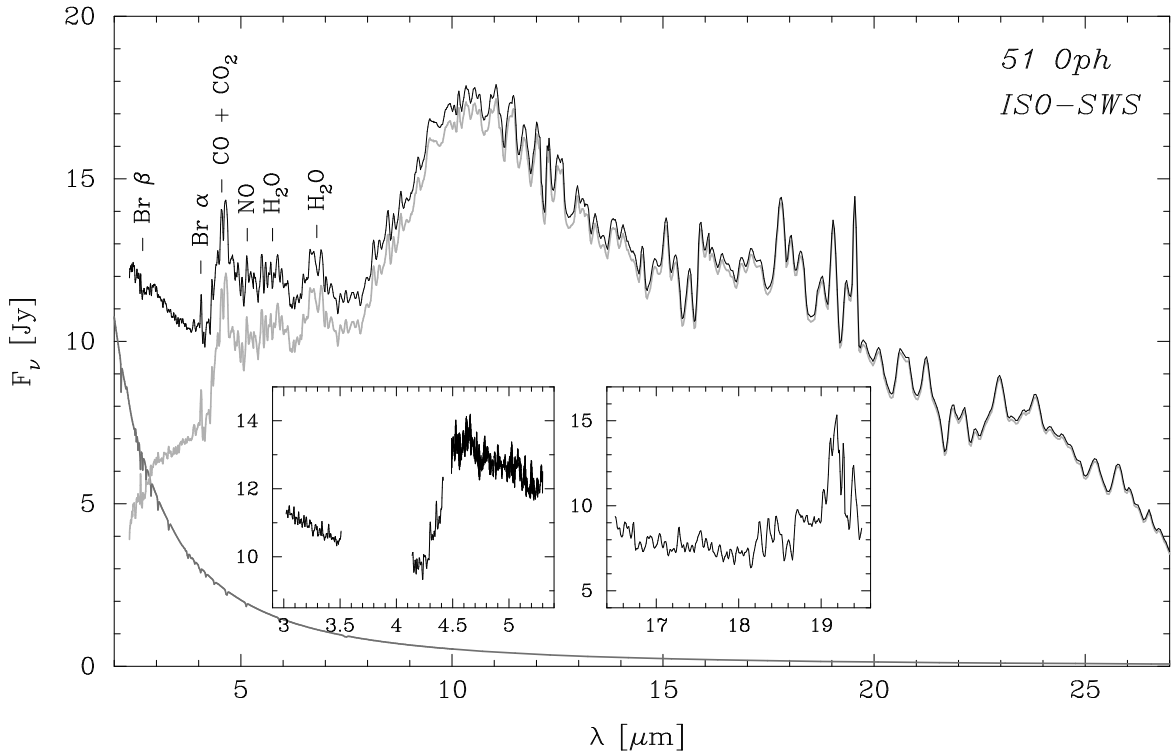}}
\caption[]{ISO-SWS spectra of 51~Oph. The main figure shows 
the AOT 01 spectrum, whereas the insets show the AOT 06 data. 
The grey lines show a Kurucz (1991) model for the A0 
primary (dark) and the spectrum (light) obtained by subtracting 
this photosphere model from the observed spectrum.}
\end{figure*}

However, the analogy with the $\beta$~Pic system is not complete.
51~Oph possesses strong H$\alpha$ emission, absent in $\beta$~Pic. The
shape of the spectral energy distribution suggests that only very few
large dust grains are present, unusual for a protoplanetary disk, but
in agreement with what is found in Vega-type systems.  However, the
magnitude and temperature of the infrared emission is more reminiscent
of disks surrounding younger stars. These indicators, together with
its position to the right of the main-sequence, lead some authors to
consider 51~Oph to be related to the Herbig Ae/Be stars (e.g. van den
Ancker et al. 1998; Meeus et al. 2001).

In this {\it letter} we will re-analyze archive ISO spectroscopy of
51~Oph. We will show that its 4--8~$\mu$m spectrum, not shown by
Waelkens et al. (1996), reveals the presence of hot circumstellar
molecular gas. This is highly unexpected for a proto-planetary or
debris disk system. We investigate several different explanations for
the remarkable composition of the circumstellar material in 51~Oph and
comment on those.

\section{Observational data}
ISO observed 51~Oph with the Short Wavelength Spectrometer (SWS; de
Graauw et al. 1996) in mode ``AOT 01'' (full spectral scan) and twice
in mode ``AOT06'' (deeper scans over a limited wavelength range). In
addition, a Long Wavelength Spectrometer (LWS; Clegg et al.  1996)
full grating scan (``AOT L01'') was taken.
The data were reduced in a standard fashion using calibration files
corresponding to ISO off-line processing software (OLP) version 7.02,
after which they were corrected for remaining fringing and glitches.
The resulting SWS spectra are shown in Fig.~1.

The spectra observed by ISO are in good agreement with the ground-based 
$N$-band spectroscopy of 51~Oph by Fajardo-Acosta et al. (1993), 
Walker \& Butner (1995) and Sylvester et al. (1996). However, the 
more complete wavelength coverage, and the higher spectral resolution 
and signal to noise reveal a wealth of structure hidden in the 
ground-based data. Shortward of 4.1~$\mu$m, the spectrum consists 
of a smooth continuum, dominated by the photosphere of the central 
star. A broad absorption due to Br$\beta$ can 
be seen around 2.63~$\mu$m. The Br$\alpha$ line at 4.05~$\mu$m is 
filled in with emission (line flux $2.5 \times 10^{-15}$~W~m$^{-2}$), 
confirming the emission-line classification of 51~Oph.

A number of strong spectral features due to gas-phase molecules are present 
in the 4.1--7.5~$\mu$m spectral range. We recognise the 4.2--5.4~$\mu$m 
fundamental vibritional band ($v$ = 1--0) of CO, the P and R branches 
of the $\nu_2$ bending mode of H$_2$O in the 5.2--7.5~$\mu$m range, 
the 4.1--4.4~$\mu$m $\nu_3$ band of CO$_2$ and the fundamental 
vibritional band of NO ($\Delta v$ = 1) around 5.1--5.9~$\mu$m. 
To further quantify the gas-phase emission lines, 
we have compared the continuum-subtracted 51~Oph spectrum to a model for 
gas-phase emission using a single layer, plane parallel LTE code (Cami et
al. 2000). A satisfactory fit to our data can be obtained with
$T_{\rm mol.}$ = 750--900~K, $N$(CO) = 10$^{20}$--10$^{22}$~cm$^{-2}$.  
The best fitting model has $T_{\rm mol.}$ = 850~K, 
$N$(H$_2$O) = $4 \times 10^{18}$~cm$^{-2}$, 
$N$($^{12}$CO$_2$) = $3 \times 10^{16}$~cm$^{-2}$, 
$N$($^{12}$CO) = $3 \times 10^{21}$~cm$^{-2}$, 
$N$($^{13}$CO) = $4 \times 10^{18}$~cm$^{-2}$, 
and $N$($^{14}$NO) $1 \times 10^{18}$~cm$^{-2}$. It is shown in Fig.~2. 
We stress that this fit is not unique and that several other combinations 
of temperature and column density also give a satisfactory fit.  However, 
we can exclude a large abundance of $^{13}$CO. The presence of NO is surprising 
but is needed in order to fit the emission structure near 5--5.5~$\mu$m.
\begin{figure}[t]
\centerline{\psfig{figure=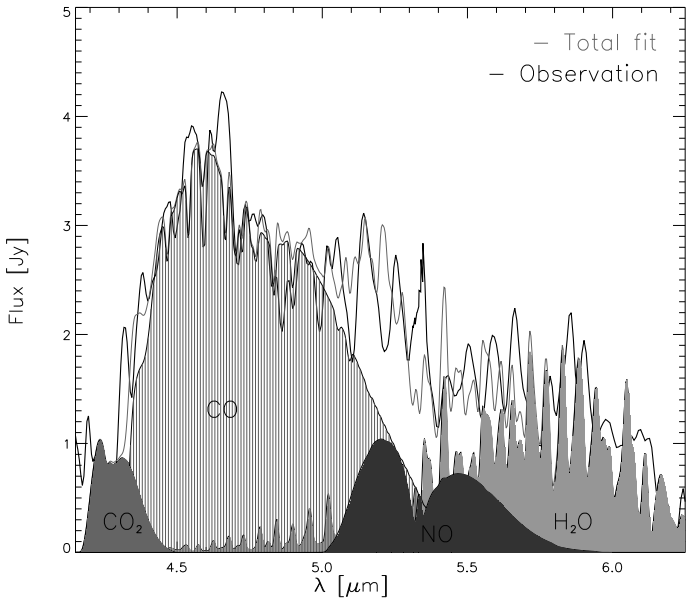,width=9.5cm,angle=90}}
\caption[]{Continuum subtracted 4--6~$\mu$m spectrum of 
51~Oph (heavy solid line) compared to a model for gas-phase emission 
of H$_2$O, CO, CO$_2$ and NO at 850~K with 
$N$(H$_2$O) = $4 \times 10^{18}$~cm$^{-2}$, 
$N$($^{12}$CO$_2$) = $3 \times 10^{16}$~cm$^{-2}$, 
$N$($^{12}$CO) = $3 \times 10^{21}$~cm$^{-2}$, 
$N$($^{13}$CO) = $4 \times 10^{18}$~cm$^{-2}$, 
and $N$($^{14}$NO) $1 \times 10^{18}$~cm$^{-2}$ (grey line). 
Contributions of the different 
gas-phase components are indicated by the shaded area.}
\end{figure}

Longward of 8~$\mu$m, the spectrum is dominated by solid-state emission 
from O-rich dust. A strong amorphous silicate feature is present, peaking 
at 10.3~$\mu$m. As was already noted by Fajardo-Acosta et al. (1993) and 
Sylvester et al. (1996), the long-wavelength shoulder of this feature consists 
of an almost linear part extending up to 15~$\mu$m. We note that this 
behaviour is incompatible with a purely silicate origin of this feature.  
A secondary feature, peaking around 11.5~$\mu$m and with a width of 
$\sim$ 3~$\mu$m must be present as well to be able to explain the shape of 
the 8--15~$\mu$m emission complex. A very broad 10 $\mu$m~band is also 
seen in the enigmatic object $\eta$ Car (Morris et al., in preparation).

A second broad bump around 18~$\mu$m is present and may be due to 
amorphous silicates. In the SWS AOT 06 spectrum a narrower (FWHM $\sim$ 
0.4~$\mu$m) peak at 19.2~$\mu$m, similar to that seen in stars with 
emission due to crystalline silicates (e.g. Molster et al. 1999), is 
superimposed on the broad amorphous silicate feature. 
51~Oph was not detected in the LWS scan.
%

\section{Discussion and Conclusions}
The presence of spectral features in 51~Oph due to hot gas-phase
molecules is highly surprising. None of the 47 Herbig Ae/Be stars for
which material is present in the ISO data archive shows a 4--8~$\mu$m
emission complex resembling the one found in 51~Oph.  However, such a
complex due to emission from hot molecular gas is commonly found in
the ISO spectra of O-rich evolved stars on or near the Asymptotic
Giant Branch (AGB; e.g. Yamamura et al. 1999, Sylvester et al. 1999).

A further inspection of the ISO data archives confirms our suspicion
that the spectrum of 51~Oph is highly unusual for a protoplanetary
system: none of the young stars for which spectroscopic data is
present shows the 11.5~$\mu$m shoulder to the amorphous silicate
feature found in 51~Oph, including the isolated Herbig Ae/Be stars, of
which 51~Oph is believed to be a member (Meeus et al. 2001; Bouwman et
al. 2001).  However, comparison of the observed 10~$\mu$m profile with
archive ISO-SWS data of the symbiotic star V835~Cen (Fig.~3) shows 
that the 10--15~$\mu$m complex in 51~Oph is
similar to the one found in this highly evolved system. In fact, the
presence of a long-wavelength shoulder to the 10~$\mu$m feature is a
common feature in the spectra of O-rich AGB objects (e.g. Tielens et
al. 1998). It is not observed in other types of stars.
\begin{figure}[t]
\centerline{\psfig{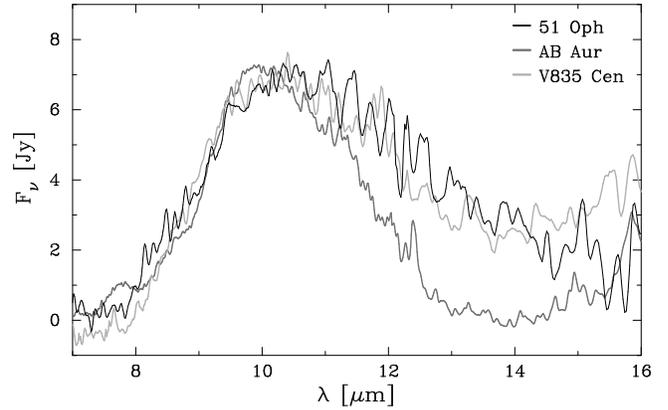}}
\caption[]{Continuum subtracted 7--16~$\mu$m spectrum of 
51~Oph (black solid line) compared to continuum-subtracted ISO 
SWS spectra of the Herbig Ae star AB Aur (van den Ancker et al. 2000; 
dark grey curve) and that of the symbiotic star V835~Cen 
(Mikolajewska et al. 1997; light grey line).}
\end{figure}

The question arises what is the origin of the circumstellar gas and
dust in 51~Oph. In view of the peculiar molecular spectrum (indicative
of a large column of hot gas) and the deviating 10~$\mu$m silicate
band, as well as the sharply dropping mid-IR to millimeter continuum
(pointing to a lack of large, cold grains typical for proto-planetary
disks) it is interesting to consider the possibility that the
circumstellar dust in 51~Oph is ``fresh'', i.e. recently formed rather
than accreted from the interstellar medium and processed in a
proto-planetary disk.

One possible explanation for the presence of hot gas and newly 
formed dust in 51~Oph would be to assume that 51~Oph is a highly  
evolved rather than a relatively young system. This assumption 
agrees well with the above main-sequence position of 51~Oph 
in the HRD.  The optical 
spectrum of 51~Oph, as well as its luminosity, show that a 
single post-main sequence star can not have created the large amount 
of circumstellar material we observe.  However, the presence of 
a cool post-main sequence companion may explain the observed 
properties of the system.  We note that a companion to 51~Oph 
may have been detected: Buscombe (1963) found the radial 
velocity of 51~Oph to be variable, which could be due to the 
reflex motion of the companion. 

By comparing the optically visible A0 star's $T_{\rm eff}$ of 
10,000~K and $L/L_{\odot}$ of 260$^{+60}_{-50}$~L$_\odot$ 
with the post-main sequence evolutionary tracks by Schaller et al. 
(1992), we derive a mass of 3.8~M$_\odot$ and an age of 
$2 \times 10^8$ years for the primary.  Using the ISO data 
and the Hipparcos distance of 131$^{+17}_{-13}$~pc towards 
51~Oph we compute an infrared excess of 7.1$^{+4.4}_{-2.4}$~L$_\odot$ 
for 51~Oph.  Assuming that both stars in our putative binary 
system formed at the same time, a lower mass companion 
should stil be on the main-sequence, limiting its mass to 
$<$ 2~M$_\odot$ if its luminosity is not to exceed the 
observed infrared excess.  This lower mass for the more 
evolved star in the system is at odds with the mass derived 
for 51~Oph itself, unless significant interaction between 
both stars occurred.  In addition, it is highly unlikely 
that an evolved star of less than 7~L$_{\odot}$ shows such 
a large amount of circumstellar matter. 

Another possibility would be to assume that the companion 
is more massive than the optically visible A0 star.  Any 
scenario in which the companion would be a red giant or 
AGB star of comparable mass to the primary is impossible, 
since such a star would have a luminosity that greatly 
exceeds 7~L$_{\odot}$.  However, if the companion would 
be sufficiently massive ($\ga$ 7~M$_\odot$), it would 
already have evolved to become a massive white dwarf of 
very low luminosity. In this scenario, the material observed 
around 51~Oph today could be caught in remnants of material 
from the AGB wind of its companion.  However, the lack of 
a significant amount of $^{13}$CO in the composition of 
the circumstellar material argues against this scenario. 
We conclude that the presence of an evolved companion in 51~Oph 
is unlikely.

The rapidly rotating ($v \sin i$ = 270~km~s$^{-1}$; Dunkin et
al. 1997), emission-line character of 51~Oph has led to an association
with the Be stars, who share many of the properties of
51~Oph. However, as pointed out by Waters et al. (1988),
circumstellar dust is highly unusual in Be stars. The presence of
double-peaked H$\alpha$ emission in 51~Oph does indicate the presence
of a gaseous disk, and the width of the line suggests that this gas is
close to the star. It is reasonable to assume that the dust is also
located in this disk, at a larger distance from the star. In Be stars,
the disk can either be the result of (time-variable) mass loss, or
from mass accretion from an evolved companion. However, as pointed out
above, the presence of an evolved cool companion is difficult to
understand given the luminosity of the dust/gas emission in the
infrared. If one accepts that the circumstellar material is due to a
recent period of high mass loss from the Be star into a disk, the
presence of dust suggests that the amount of gas ejected must have
been high enough to shield the UV radiation field of the central star
to allow for dust formation. We have fitted the ISO-SWS spectrum using
the dust radiative transfer code MODUST (Bouwman \& de Koter, in
preparation) and find a dust mass of $4 \times 10^{-9}$~M$_{\odot}$. 
This is a factor of 5 higher than the (distance-corrected) value given 
by Fajardo-Acosta et al. (1993). In this scenario the late-main 
sequence nature of 51~Oph may be related to the ejection of a large 
amount of mass, since rapidly rotating early type stars are believed 
to move closer to their breakup velocity as they evolve 
(e.g. Langer et al. 1999; Langer 2000).

A more exotic possibility to explain the properties of 51~Oph
would be to infer that we are seeing the aftermath of a recent
event such as the collision of two gas-rich planets or the
accretion of a solid body as the star increases its size at the
end of its main-sequence life. New dust could form at the site of
the evaporation of the solid body, explaining both the high column
of hot gas, the apparent small dust particle sizes and the
composition of the silicate dust.

As was already pointed out by previous authors (e.g. Herbig 1994) the
problem of separating young stars from their evolved counterparts is
non trivial. The case of 51~Oph presented in this {\it letter} may
serve once more to illustrate this difficulty.  Even with the
extensive data present on this enigmatic object, its evolutionary
status remains unclear. New data that is able to resolve the system
spatially, as well as a more thorough investigation of the suspected
radial velocity variations, may be needed to shed more light on its
nature.

\acknowledgements{The authors would like to thank Dr. Frank Molster 
for useful discussions regarding the nature of 51~Oph. We thank
Dr. Alex de Koter and Drs. Jeroen Bouwman for their help in the
use of MODUST.  We are also grateful to the referee, Mike Barlow, 
for many useful comments.  
This research has made use of the Simbad data base, operated at 
CDS, Strasbourg, France. LBFMW acknowledges financial support from 
an NWO \emph{Pionier} grant.}


\begin{thebibliography}{}
\bibitem{}
Bouwman J., et al., 2001, A\&A, in press
\bibitem{}
Buscombe W., 1963, MNRAS 126, 29
\bibitem{}
Cami J., Yamamura I., de Jong T., Tielens A.G.G.M., Justtanont K., 
 Waters L.B.F.M., 2000, A\&A 360, 562
\bibitem{}
Clegg P.E., Ade P.A.R., Armand C., et al., 1996, A\&A 315, L38
\bibitem{}
Cot\'e J., Waters L.B.F.M., 1987, A\&A 176, 93
\bibitem{}
de Graauw Th., Haser L.N., Beintema D.A., et al., 1996, 
 A\&A 315, L49
\bibitem{}
Dunkin S.K., Barlow M.J., Ryan S.G., 1997, MNRAS 286, 604
\bibitem{}
Fajardo-Acosta S.B., Telesco C.M., Knacke R.F., 1993, ApJ 417, L33
\bibitem{}
Grady C.A., Silvis J.M.S., 1993, ApJ 402, L61
\bibitem{}
Gray R.O., Corbally C.J., 1998, AJ 116, 2530
\bibitem{}
Herbig G.H., 1994, 
 ASP Conf. Series 62, p. 3
\bibitem{}
Jaschek C., Jaschek M., 1992, A\&AS 95, 535
\bibitem{}
Kessler M.F., Steinz J.A., Anderegg M.E., et al., 1996, A\&A 315, L27
\bibitem{}
Kurucz R.L., 1991, in ``Stellar atmospheres--Beyond classical models''
 (eds. A.G. Davis Philip, A.R. Upgren, K.A. Janes), L. Davis press, 
 Schenectady, New York, p. 441
\bibitem{}
Lagage P.O., Pantin E., 1994, Experimental Astron. 3, 57
\bibitem{}
Langer N., 2000, Science 287, 2430
\bibitem{}
Langer N., Heger A., Wellstein S., Herwig F., 1999, A\&A 346, L37
\bibitem{}
Lecavelier des Etangs A., Vidal-Madjar A., Backman D.E., et al., 
 1997, A\&A 321, L39
\bibitem{}
Lynch D.K., Russell R.W., Hackwell J.A., Fajardo-Acosta S., Knacke R., 
 Hanner M., Telesco C., 1994, AAS 184, 1607
\bibitem{}
Meeus G., Waters L.B.F.M., Bouwman J., van den Ancker M.E., 
 Waelkens C., Malfait K., 2001, A\&A, in press
\bibitem{}
Merkle F., Gehring G., Rigaut F., Kern P., Gigan P., Rousset G., 
 Boyer C., 1990, Messenger 60, 9
\bibitem{}
Mikolajewska J., Acker A., Stenholm B., 1997, A\&A 327, 191
\bibitem{}
Molster F.J., Waters L.B.F.M., Trams N.R., et al., 1999, A\&A 350, 163
\bibitem{}
Schaller G., Schaerer D., Meynet G., Maeder A., 1992, A\&AS 96, 269
\bibitem{}
Slettebak A., 1982, ApJS 50, 55
\bibitem{}
Sylvester R.J., Skinner C.J., Barlow M.J., Mannings V., 1996, 
 MNRAS 279, 915
\bibitem{}
Sylvester R.J., Kemper F., Barlow M.J., de Jong T., Waters L.B.F.M., 
 Tielens A.G.G.M., Omont A., 1999, A\&A 352, 587
\bibitem{}
Tielens A.G.G.M., Waters L.B.F.M., Molster F.J., Justtanont K., 1998, 
 Ap\&SS 255, 415
\bibitem{}
van den Ancker M.E., Bouwman J., Wesselius P.R., Waters L.B.F.M., 
 Dougherty S.M., van Dishoeck E.F., 2000, A\&A 357, 325
\bibitem{}
van den Ancker M.E., de Winter D., Tjin A Djie H.R.E., 1998, 
 A\&A 330, 145
\bibitem{}
van der Veen W.E.C.J., Habing H.J., 1988, A\&A 194, 125
\bibitem{}
Waelkens C., Waters L.B.F.M., de Graauw M.S., et al., 1996, A\&A 315, L245
\bibitem{}
Walker H.J., Butner H.M., 1995, Ap\&SS 224, 389
\bibitem{}
Waters L.B.F.M., Cot\'e J., Geballe T.R., 1988, A\&A 203, 348
\bibitem{}
Yamamura I., de Jong T., Cami J., 1999, A\&A 348, L55
\end{thebibliography}
\end{document}